# Deep Neural Network: An Efficient and Optimized Machine Learning Paradigm for Reducing Genome Sequencing Error.

Dr. Ferdinand Kartriku [#1], Robert Sowah [*2], Charles Saah [#3]

[#1] Senior Lecturer Computers Science Department
[#2] HoD Computers Engineering Department
[#3] Computers Engineering Department
university of Ghana – Legon.
[1] fkatsriku@ug.edu.gh
[2] rasowah@ug.edu.gh
[3] cscharles393@gmail.com

*Abstract*— Genomic data is used in many fields but, it has become known that most of the platforms used in the genome sequencing process produce significant errors. This means that the analysis and inferences generated from these data, may have some errors that need to be corrected. On the two main types (substitution and indels) of genome errors, our work focused on correcting errors emanating from indels. A deep learning approach was used to correct the errors in sequencing the chosen dataset.

**Keywords —** *genome sequencing; error correction; deep learning; indels;*

## I. INTRODUCTION

Since the time of Sanger, many genome sequencing projects have emerged. All the projects are geared towards improving the genome sequencing process. Each sequencing project introduces some level of error or variants in the sequenced data. This is mainly due the underlying methods or mechanism that the sequencing process undergoes [1]. As the sequencing processes grow, so does the errors introduced based on the sequencing process [2]. Distinguishing between variants that emanate from the sequencing process is technologically and computationally challenging. Research has established that, the errors can be categorized into two main domains [3]. That is, errors due to substitution of nucleotide, and what has become known as indel, that is insertion or deletion errors. The application of the advancement in data science, mathematics and computer science in biology has brought on board a myriad of attempts aimed at solving this problem. The ushering in of the next generation sequencing process (NGS) which was geared towards improving and simplifying the sequencing process also introduced errors in the sequenced data [1]. It has also been established that not only does the sequencing process generate errors but also, the choice of data used in the sequencing process can contribute to the underlying errors.[4][2].

Most of the error corrections methods have been aimed at substitution errors emanates from the Illumina sequencing platform [5][1][6][7]. The main purpose of this work is to reduce insertion and deletion errors by designing and optimizing a deep convolution neural network that drastically reduce genome sequencing error and also reduce computational time for sequencing while using minimal computer resources. Next we are going to review literature on genome sequencing error corrections, that will be followed by our methodology which will usher in our results, then discussion and conclusion.

## II. LITERATURE REVIEW

The quest to correct sequencing errors increase tremendously from the detection of variation in the human DNA and sequencing reads from RNA [8][9]. However most of the initial error correction processes focused on remedying substitution errors as majority of them focused on correcting errors generated by the Illumina sequencer [5]. Crosstalk sequencing error from the Illumina sequencing process where the dye used, exhibited overlapping signal strength characteristics leading to the misinterpretation of nucleotides such as A for C and G for T is known to contribute immensely to substitution errors [10][11]. Again the continuous k-mer generation from nucleotide also leads the replication of an error throughout the sequencing process [5][12], thus bloating the size of the error in the sequencing process. Inverted sequencing repeats of nucleotides such as GGC which is known as dephasing has also been identifies as a source of sequencing errors besides location specific alignment emanating from k-mer of fixed read length [13].

Platforms such as Roche's 454, Ion Torrent are known to introduce indels in the sequencing process [14] [15]. Reference sequence error correction are heavy on computer memory usage and it is time consuming [1]. Sequencing error is unavoidable because of the processes used in sequencing genomic data, however the ability to identify and correct them, if not completely eliminate them is paramount [16]. Several works have been done in the arena of genome sequencing error correction. There are two main approaches in genome sequencing error correction, that is using a reference genome and not using a reference genome. The reference approach compares the sequenced data with a known (reference) sequence of the same genome data type. The challenge with this approach is that in certain situations, there are no reference genome available for use [1]. Several works have been done on sequencing with and without a reference genome [17] [18] [19] [20]. It therefore indicative to say that sequencing with reference genome outperforms those without a reference genome.

In correcting substitution errors, [15] [6] [21] used the k-spectrum approach where the probability of a k-mer occurring a certain number of times were classified as solid and those outside the specified number of times were classified as in-solid. The weighted sum of solid and in-solid were then computed and a histogram plotted. The solid was said to follow a blend of Gaussian and zeta distribution while the in-solid followed a Gamma distribution [2]. Quality values representing the number of times each occurred were computed and proposed that the sequencing error followed the probability distribution of the quality values. Further research by Dohn J. C, Lottaz C, et al [16] showed that the assertion was not the necessary the case

Suffix tree or array based methods were also used to correct insertion and deletion errors [3] [22]. This was done by treating k-mers as forming tree or array data structure. In an iteration process, if a k-mer is considered as an error, it is compared with the children of the root in the structure and any insertion or deletion errors are corrected

### III. METHODOLOGY

A deep convolutional neural network architecture which uses sliding window emanating from learned filters to automatically detect patterns at various locations was designed. Our model consists of three hidden layers, each hidden layer consist of convolutional network, ReLU activation function, maxpool layer which reduces the size of the input volume for the next layer. A flatten layer then converts the maxpool featured map into a column vector for the fully connected layer. A dropout layer is then used to trim the network to prevent overfitting. The output of the dropout layer is then passed through another fully connected layer before passing it through the softmax probability function to predict the output data. The data NA12878, taken from the National Centre for Biotechnology Information (NBCI) was divided into training, validation and testing respectively, using the 80%, 10% and 10% ratio. We used a one-hot encoding scheme where the nucleotide bases A, C, T and G were respectively encoded as [0 1 0 0], [1 0 0 0], [0 0 1 0] and [0 0 0 1]. The network architecture is shown in figure 1.

Instead of correcting errors in single reads, we used a consensus based approach where we built consensus of multiple reads and focused on generating underlying DNA. Figure 2 depict the convolutional network connections.

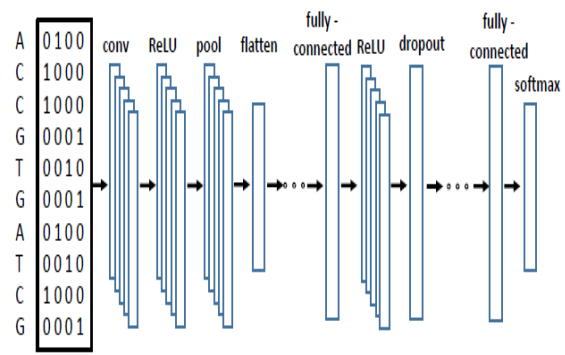

**Fig: 1. Convolutional Neural Network using the one-hot encoded scheme as input data for training the network and the softmax function to predict the output**

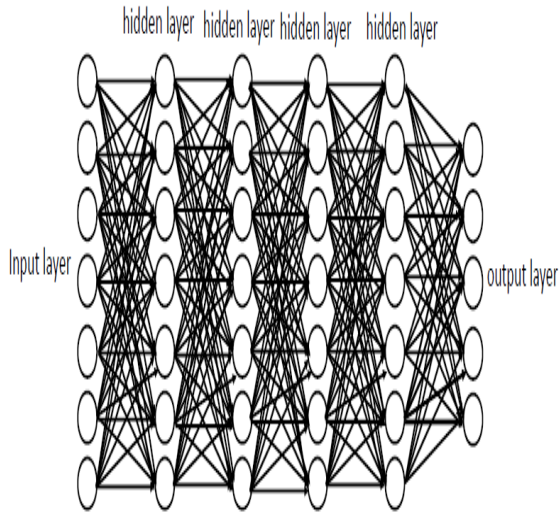

**Fig 2: Fully connected convolutional neural network with four hidden layers.**

The network was then trained and validated please see figure 3 below. The validation process between epoch 0 and 5 seemed good but took a divergent tangent after epoch 5 and did not recover even after epoch 50.

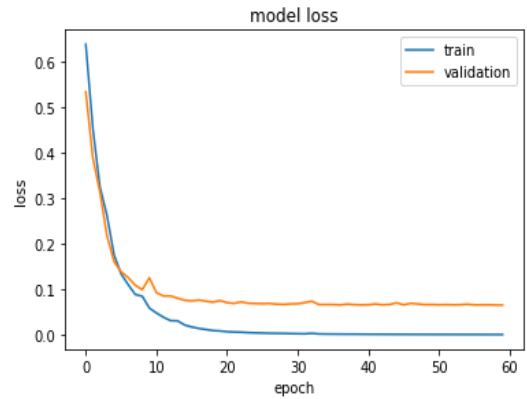

**Fig 3: Initial training and validation of the network showing divergence in the validation after epoch 5.**

For the network to perform better on the testing data, the validation process has to be 100% based on the training data used. The hyper parameters of the network was then tweaked to improve on the training and validation process. The training and validation process improved tremendously see figure 4(a and b).

## IV. RESULTS

Figure 6 shows that the network has a high accuracy of 99.2% in sequencing the data. Figure 4 *b* also show a high validation of the trained dataset with close to zero loss after epoch 20.

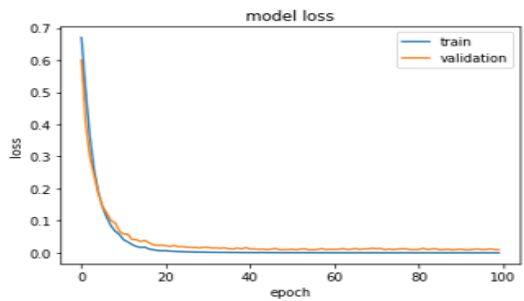
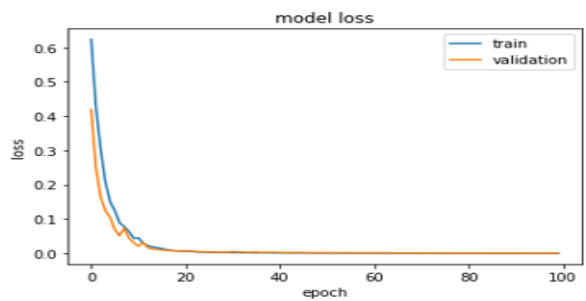

**Fig 4: Improved network validation process after tweaking of hyper parameters.**

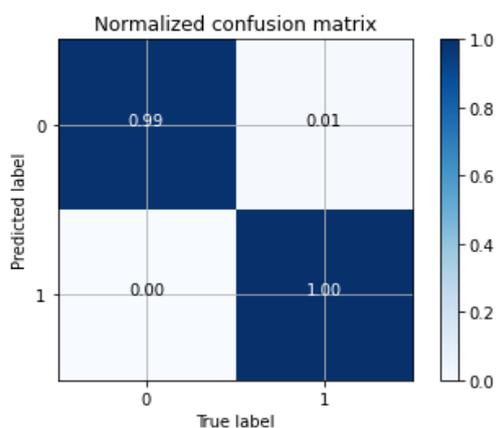

**Fig 5: Normalized Confusion Matrix.**

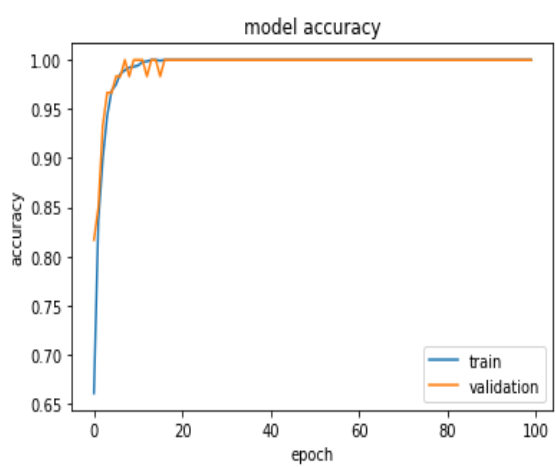

**Fig 6: Network accuracy diagram.**

## V. DISCUSSION

The deep convolutional neural network through consensus sequencing has been able reduce insertion and deleting error to the barest minimum. This is showcased in figure 4b where the system validated all the training datasets with zero loss. The normalized confusion matrix in figure 5, displayed a performance of 99%. This was achieved after epoch 40 and the network performance in figure 6 remained stable through epoch 100. This demonstrates the resilience of the network in predicting the genome given an input data.

The experiment was conducted using Hewlett packed pavilion core i5 laptop, with 12GB RAM and 1 Terabyte hard disk. The process run smoothly without any hindrances to the functionalities of the computer and applications that run concurrently.

Compared to similar experiments by [3] [17] our network performed better. We must say that different datasets were used in our experiment and theirs.

The choice of deep CNN which has the capacity to apply learning features to input dataset as it does in image recognition and natural language processing helped in the network performance. This is mainly because the network adds weights and biases during the feedforward process and automatically adjust the weights and biases during the backpropagation process thus improving on the learning process.

## VI. CONCLUSION

we have been able to demonstrate that genome sequencing error correction particularly indels can be achieved without compromising on system resources and computational prowess. Though the accuracy of 99.2% is near perfect, we will like to try other architectures using the same or different dataset to improve on the network performance. If the new architecture works successfully, it will be extended to correcting substitution errors